\documentclass{webofc}
\usepackage[varg]{txfonts}
\usepackage[utf8]{inputenc}
\usepackage{hyperref}
\hypersetup{
    colorlinks=true,
    linkcolor=blue,
    filecolor=blue,      
    urlcolor=cyan,
    }
\begin{document}
\title{Holographic dense QCD in the Veneziano limit}
%
%

\author{
\firstname{Matti} \lastname{J\"arvinen}\inst{1,2}
}

\institute{Asia Pacific Center for Theoretical Physics, Pohang, 37673, Korea
\and
Department of Physics, Pohang University of Science and Technology, Pohang, 37673, Korea
}

\abstract{%
Solving the properties of dense QCD matter is an extremely challenging problem because standard theoretical tools do not work at intermediate densities. The gauge/gravity duality may help to provide answers in this region. I give a brief review of recent progress in this field, focusing on the V-QCD model, which is one of the most sophisticated holographic models of QCD. I discuss predictions for the phase diagram, the equation of state, and properties of baryons. I apply these results to analyze the properties of neutron stars and to quark matter production in neutron star mergers.
}
{
\hfill APCTP Pre2022 - 025
}
\maketitle
\section{Introduction} 
\label{intro}
New and ongoing observations of neutron stars provide more and more information on their properties, starting from basic observables such as their masses and radii. These observations provide, in turn, information on characteristics of dense QCD matter found inside the stars~\cite{Annala:2017tqz,Annala:2019puf,Altiparmak:2022bke}. 
In particular, the recent observation of gravitational and electromagnetic events the neutron star merger event GW170817~\cite{LIGOScientific:2017zic} sets tight bounds on the QCD equation of state (EOS). In addition, new kind of heavy ion collision experiments, such as the ongoing beam energy scan at RHIC and planned experiments at FAIR and NICA, will give complementary information on dense matter at finite temperature.

Therefore, it is timely to improve the status of theoretical knowledge of dense QCD~\cite{Brambilla:2014jmp}. This is a difficult task as lattice QCD computations are currently limited to low densities due to the sign problem~\cite{deForcrand:2010ys}. While progress is being made with effective methods at low density (see, e.g.,~\cite{Drischler:2017wtt}) and perturbative QCD at high density (see, e.g,~\cite{Gorda:2021znl}), uncertainties at intermediate densities are large. Consequently, modeling can provide important information in this region, which is known to be relevant for neutron stars.

Using the gauge/gravity duality for this purpose has gained increasing interest in recent years. Several groups are making progress by using a variety of top-down holographic models (such as D3-D7~\cite{Hoyos:2016zke,BitaghsirFadafan:2019ofb} and Witten-Sakai-Sugimoto~\cite{Kovensky:2020xif,Pinkanjanarod:2020mgi,Kovensky:2021kzl} models) and by using bottom-up holography (e.g. Einstein-Maxwell~\cite{Ghoroku:2019trx,Mamani:2020pks,Ghoroku:2021fos} and hard wall~\cite{Bartolini:2022rkl} models). Here I will review progress with another bottom-up approach, the V-QCD model~\cite{Jarvinen:2011qe}. 
  

\section{The V-QCD model}

The V-QCD model is a complex bottom-up holographic model (see the recent review~\cite{Jarvinen:2021jbd}). It makes use of the Veneziano limit~\cite{Veneziano:1976wm},
\begin{equation}
 N_c \to \infty, \qquad N_f \to \infty\ , \qquad x \equiv N_f/N_c \quad \mathrm{fixed}\,.
\end{equation}
The model is
obtained by combining two building blocks:
\begin{enumerate}
 \item Improved holographic QCD (IHQCD), a model for pure Yang-Mills theory inspired by five dimensional noncritical string theory (Einstein-dilaton gravity)~\cite{Gursoy:2007cb,Gursoy:2007er}. 
 \item Method for introducing flavors and chiral symmetry breaking via Sen-like~\cite{Sen:2004nf} tachyon Dirac-Born-Infeld (DBI) action for space filling pairs of flavor branes~\cite{Bigazzi:2005md,Casero:2007ae}.
\end{enumerate}
In the Veneziano limit the flavor sector fully backreacts to the glue. 


Here I will be using the model at finite temperature and density. 
To this end, I consider a metric with a horizon, so that the thermodynamics comes from that of the planar bulk black hole~\cite{Gursoy:2008za,Alho:2012mh}. Finite baryon number density arising from free quarks is obtained through a nonzero temporal component of the Abelian gauge field in the DBI action~\cite{Alho:2013hsa,Alho:2015zua}. 

An essential ingredient of the model is that it contains several functions of the dilaton field which need to be carefully chosen in order to mimic properties of QCD. These include the dilaton potential from the glue sector as well as similar potential functions appearing in the DBI action. The asymptotics of these functions at small and large dilaton are chosen so that the model shows asymptotic freedom, discrete spectrum, linear confinement, qualitatively correct phase diagram at finite temperature and density, as well as correct behavior at finite quark mass and $\theta$-angle~\cite{Gursoy:2007cb,Gursoy:2007er,Jarvinen:2011qe,Arean:2013tja,Jarvinen:2015ofa,Arean:2016hcs,Ishii:2019gta}. The behavior of the functions at intermediate values of the dilaton can then be tuned to agree with lattice data for glueball spectra, thermodynamics of Yang-Mills~\cite{Gursoy:2009jd} and full QCD~\cite{Jokela:2018ers}, or experimental data for meson masses~\cite{Amorim:2021gat}. 
A combined, overall fit will be discussed in a future publication~\cite{Jarvinen:2022xxx}.

\begin{figure*}[tbh] 
\centering
     \includegraphics[width=0.44\textwidth]{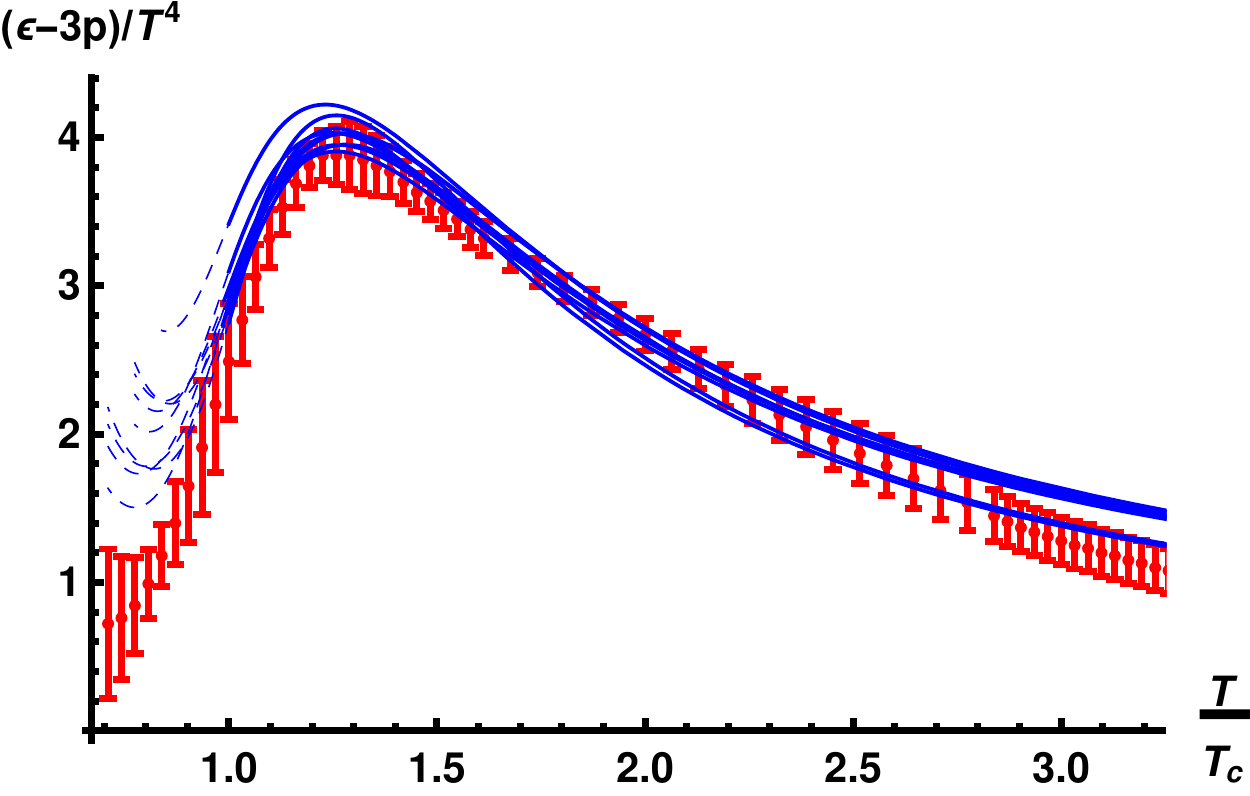}\quad
     \includegraphics[width=0.44\textwidth]{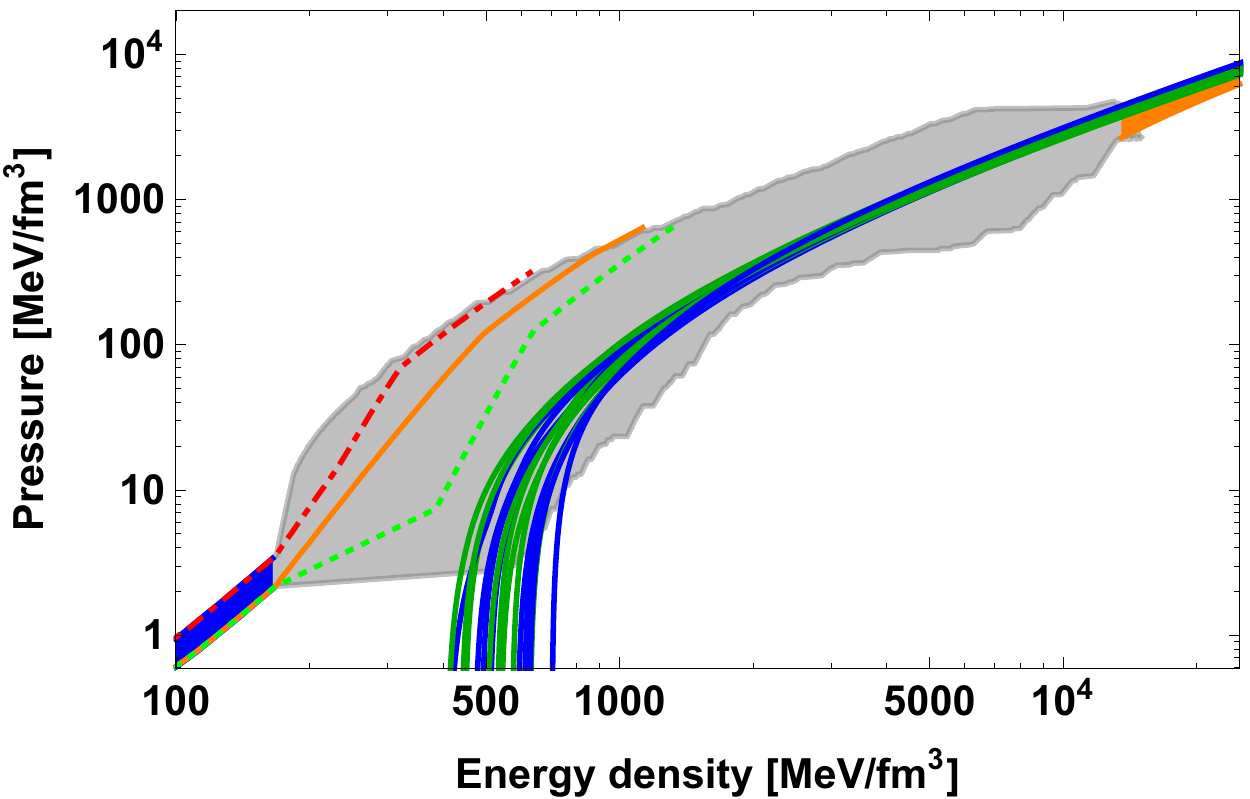}
    \caption{\small Thermodynamics in the V-QCD model. Left: Fit of the model (blue curves) to lattice data~\cite{Borsanyi:2013bia} (red dots and error bars). Right: Extrapolation of the EOS to zero temperature (blue and green curves), compared to the band of interpolated EOSs~\cite{Annala:2017llu} and to nuclear matter EOSs~\cite{Hebeler:2013nza}. Figures from~\cite{Jokela:2018ers}.}
    \label{fig:thermo}
\end{figure*}

Here I will be using the fit of~\cite{Jokela:2018ers}, which describes the QCD thermodynamics at high precision, see Fig.~\ref{fig:thermo} (left). Despite the large number of parameters in the model, this is nontrivial, because the parameter dependence is mild and all potential functions need to be simple monotonic functions with known asymptotics for the model to work properly. After fitting the lattice data to QCD thermodynamics, the EOS of (unpaired) quark matter at low temperatures and high densities is found to be feasible, and also roughly agrees with perturbative analysis at high temperatures and densities, see Fig.~\ref{fig:thermo} (right). Therefore the model of quark matter works well at essentially all relevant values of temperature and density.

\section{Applications and extensions}
 
In the past few years, V-QCD and IHQCD have been applied to physics at finite magnetic field~\cite{Gursoy:2017wzz}, in the presence of anisotropy~\cite{Gursoy:2020kjd}, and to analyze Regge physics~\cite{Ballon-Bayona:2017vlm,Amorim:2018yod,Amorim:2021ffr}. Most of the interest has however been on the applications to dense matter, which I will review here.

\subsection{Homogeneous approach to nuclear matter}

\begin{figure*}[htb]
\centering
     \includegraphics[width=0.45\textwidth]{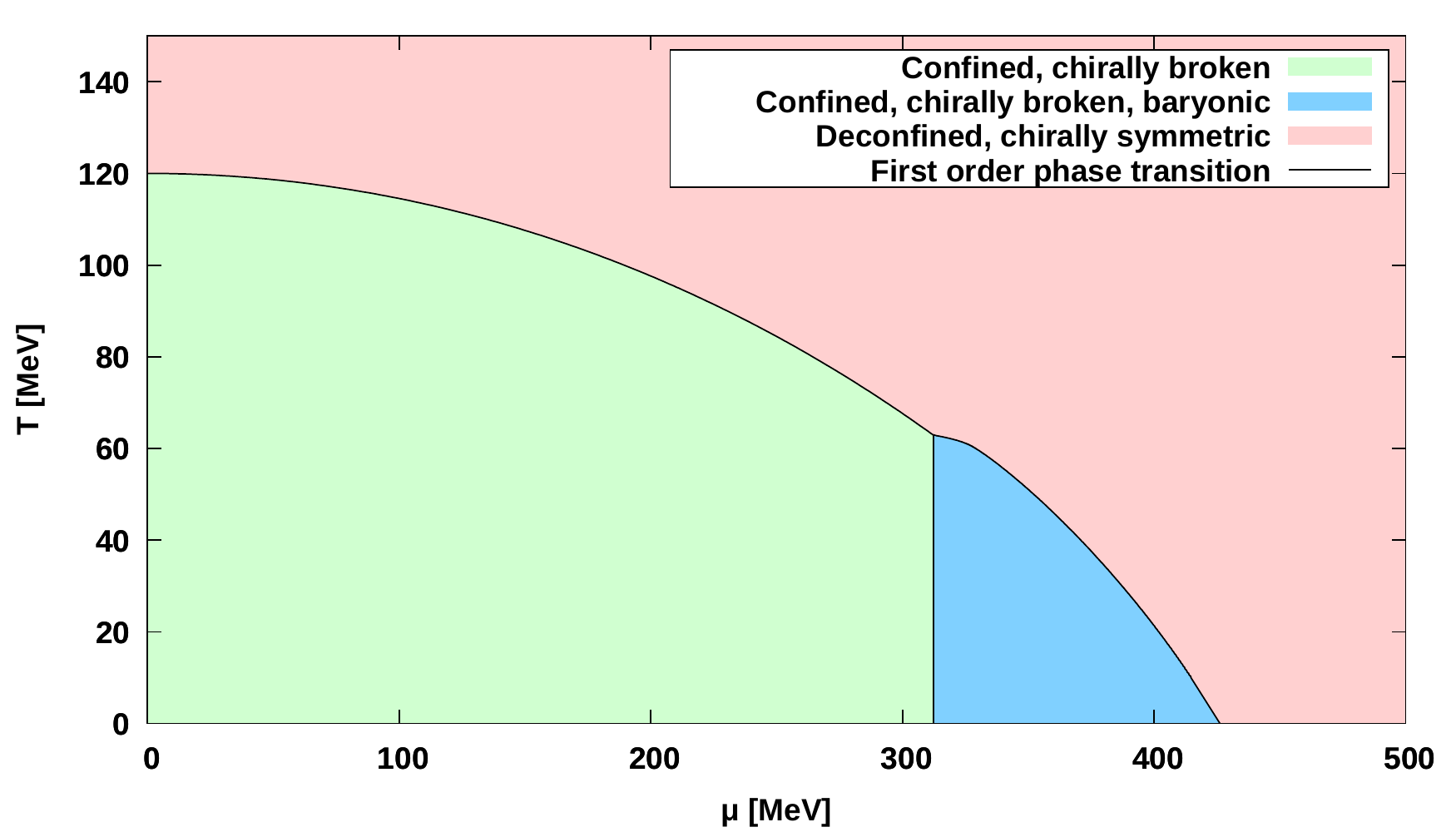}\quad
     \includegraphics[width=0.45\textwidth]{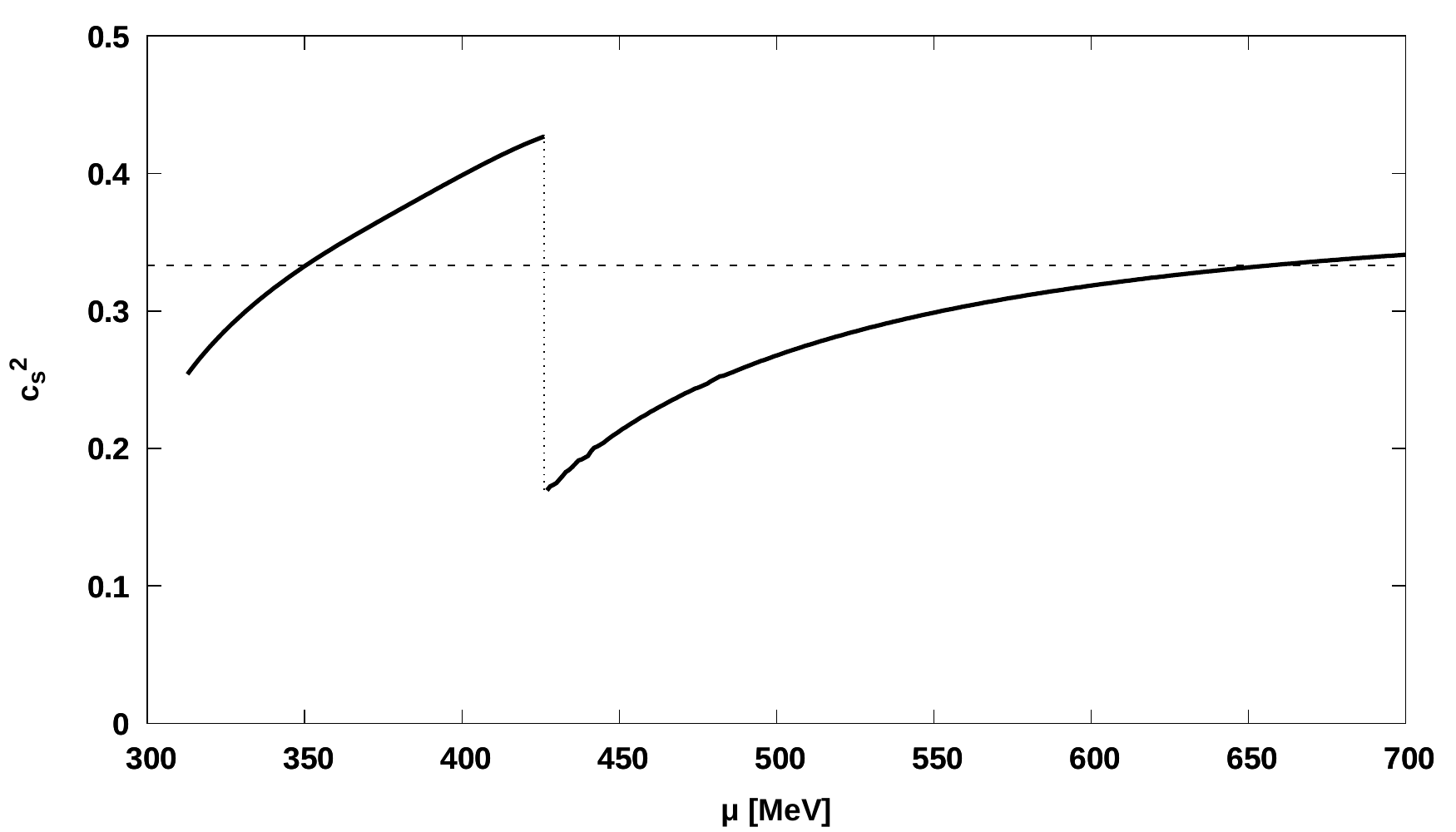}
    \caption{\small Adding homogeneous nuclear matter in the V-QCD model. Left: The phase diagram on the $(\mu,T)$-plane. Right: The speed of sound at zero temperature. Figures from~\cite{Ishii:2019gta}.}
    \label{fig:homognucl}
\end{figure*}

The V-QCD model as discussed above already includes dense quark matter, but not nuclear matter. Baryons in the holographic approach are obtained through solitonic ``instanton'' configurations of the non-Abelian gauge fields living on the D-branes. 
While it is possible to construct such solitons in V-QCD (see Sec.~\ref{sec:baryon}), 
building a dense matter out of solitons is an extremely challenging problem. Therefore 
we resort to a simple approach with a homogeneous non-Abelian gauge field~\cite{Rozali:2007rx,Li:2015uea} with two flavors,
$
 A^i = 
 h(r)\, \sigma^i \,,
$
where $i=1,2,3$ are Lorentz indices, $r$ is the holographic coordinate, 
and $\sigma^i$ are the Pauli matrices (in flavor space). 
Applying this method in V-QCD~\cite{Ishii:2019gta} leads to the phase diagram of Fig.~\ref{fig:homognucl} (left). Interestingly the speed of sound in Fig.~\ref{fig:homognucl} (right) is high in the nuclear matter phase: the EOS is stiff. This 
will help building viable models that can support highly massive neutron stars.

\subsection{(Cold) hybrid equations of state and application to neutron stars}

The V-QCD model 
however cannot produce a fully feasible EOS for use in neutron star physics, because the homogeneous nuclear matter approach fails in the crust of the stars where densities are relatively low. Indeed, for densities below the nuclear saturation density $n_s \approx 0.16\, \mathrm{fm}^{-3}$ nucleons are clearly separated, and cannot be reliably modeled by a homogeneous distribution. Our approach is simply to replace holography in this regime by ``traditional'' approaches based on effective theory. It is actually difficult to derive a feasible EOS at low densities from holography (see however~\cite{Kovensky:2021ddl}). We choose to use the popular APR~\cite{Akmal:1998cf} and HS(DD2)~\cite{Hempel:2009mc,Typel:2009sy} EOSs at low densities (i.e., $n \lesssim 1.5 n_s$) together with V-QCD at higher densities, and use them to construct families of ``hybrid'' EOSs at $T=0$~\cite{Ecker:2019xrw,Jokela:2020piw,Demircik:2021zll}.


\begin{figure*}[thb]
\centering
     \includegraphics[width=0.88\textwidth]{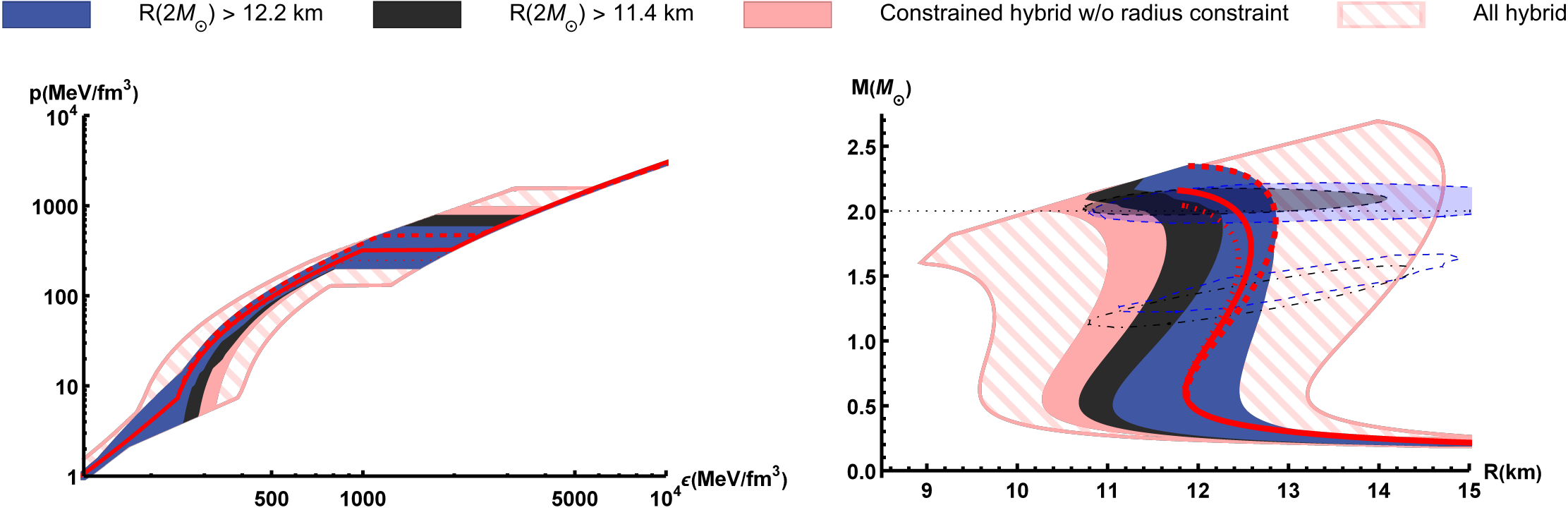}
    \caption{\small Hybrid V-QCD EOSs at $T=0$. Left: the regions spanned by allowed by the allowed EOSs as indicated by the labels. Right: The allowed regions of mass-radius curves.  Figures adapted from~\cite{Jokela:2020piw}.}
    \label{fig:hybrid}
\end{figure*}

After completing the EOSs the simplest application is to analyze the structure of neutron stars. 
Results can be confronted with observations of neutron star masses and radii, as well as the deformation of neutron stars during the merger event GW170817~\cite{Abbott:2018exr}. The mass measurements, in particular the result  $M =  2.08\pm 0.07 M_{\odot}$~\cite{Fonseca:2021wxt} for the pulsar J0740+6620 indicates that the maximum mass of neutron stars must be around $2M_\odot$ or higher. The effect of these constraints on the hybrid EOSs are demonstrated in Fig.~\ref{fig:hybrid} (left) as the pink band. We also added the constraints from NICER measurements of J0740+6620 (blue and black bands)~\cite{Miller:2021qha,Riley:2021pdl}, and the EOSs for three examples of hybrid EOSs as the red curves. The right plot shows the same for mass-radius curves of nonrotating neutron stars.

Rapidly rotating neutron stars were analyzed using the V-QCD EOSs in~\cite{Demircik:2020jkc}. This study was motivated by the gravitational wave observation GW190425~\cite{Abbott:2020khf}: a merger of a heavy ($\sim 23 M_\odot$) black hole with a lighter ($\sim 2.6 M_\odot$) companion. The lighter object falls in the ``mass gap'',  it is not clear whether it is a black hole or a neutron star. We found that our EOSs are compatible with the interpretation of this object as a neutron star, but it had to be rapidly rotating: the rotation frequency would need to be clearly higher than that of the fastest known pulsar. For the ratio of maximal masses of rotating and static neutron stars we found
\begin{equation} \label{eq:massratiofit}
    M_\mathrm{max}/M_\mathrm{TOV} = 1.227^{+0.031}_{-0.016} \,, 
\end{equation}
a number slightly higher than in models without phase transition~\cite{Breu:2016ufb}.


\subsection{Transport in dense matter}

While the EOS of QCD has large uncertainties at the densities of neutron star cores, much less is know about transport properties (see the review~\cite{Schmitt:2017efp}). Transport is important for neutron star cooling, for damping of oscillations of the stars, and is also estimated to be relevant in neutron star mergers in the most violent period, right after the merger.

\begin{figure*}[htb]
\centering
     \includegraphics[width=0.42\textwidth]{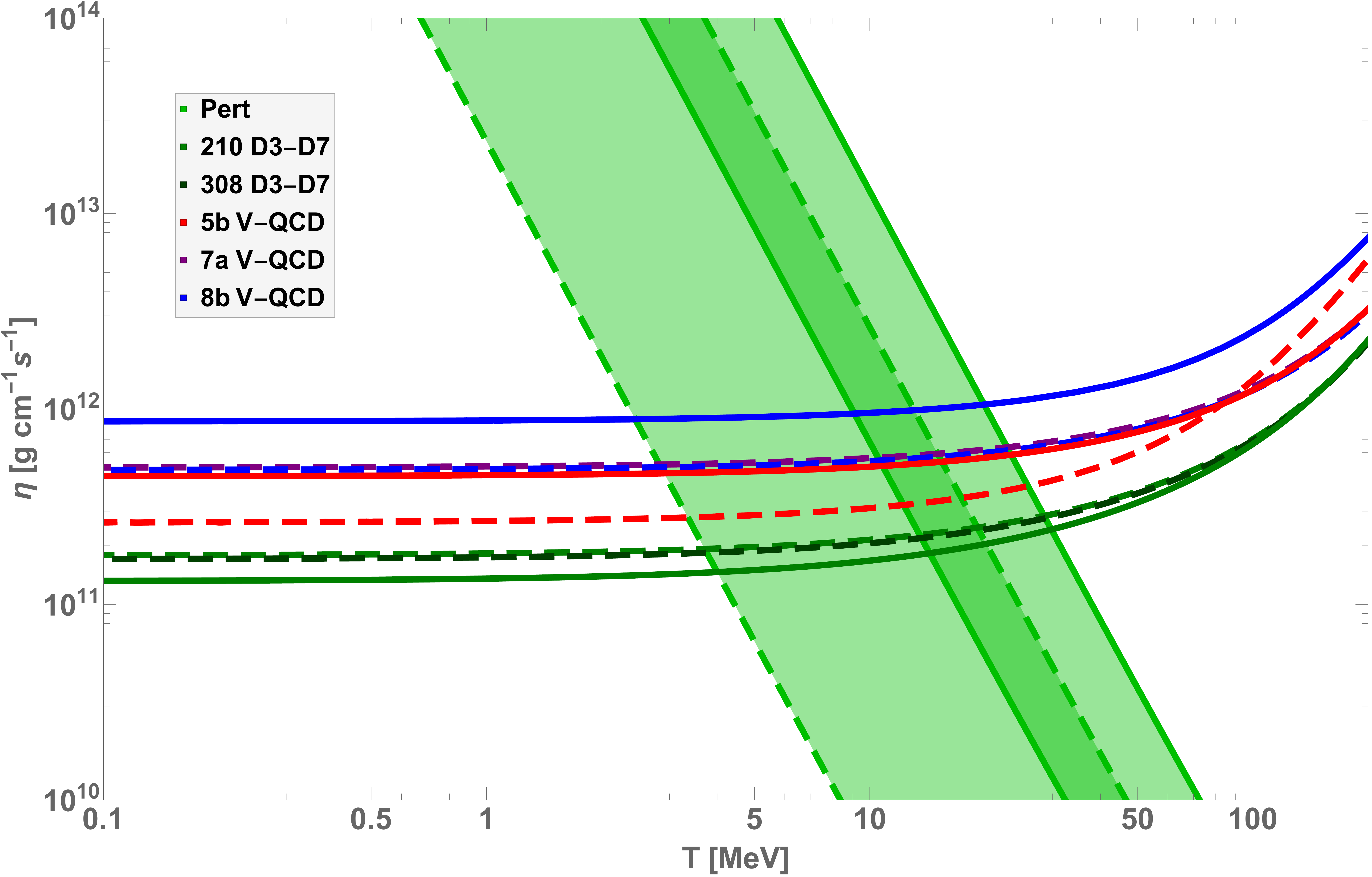}\quad
     \includegraphics[width=0.42\textwidth]{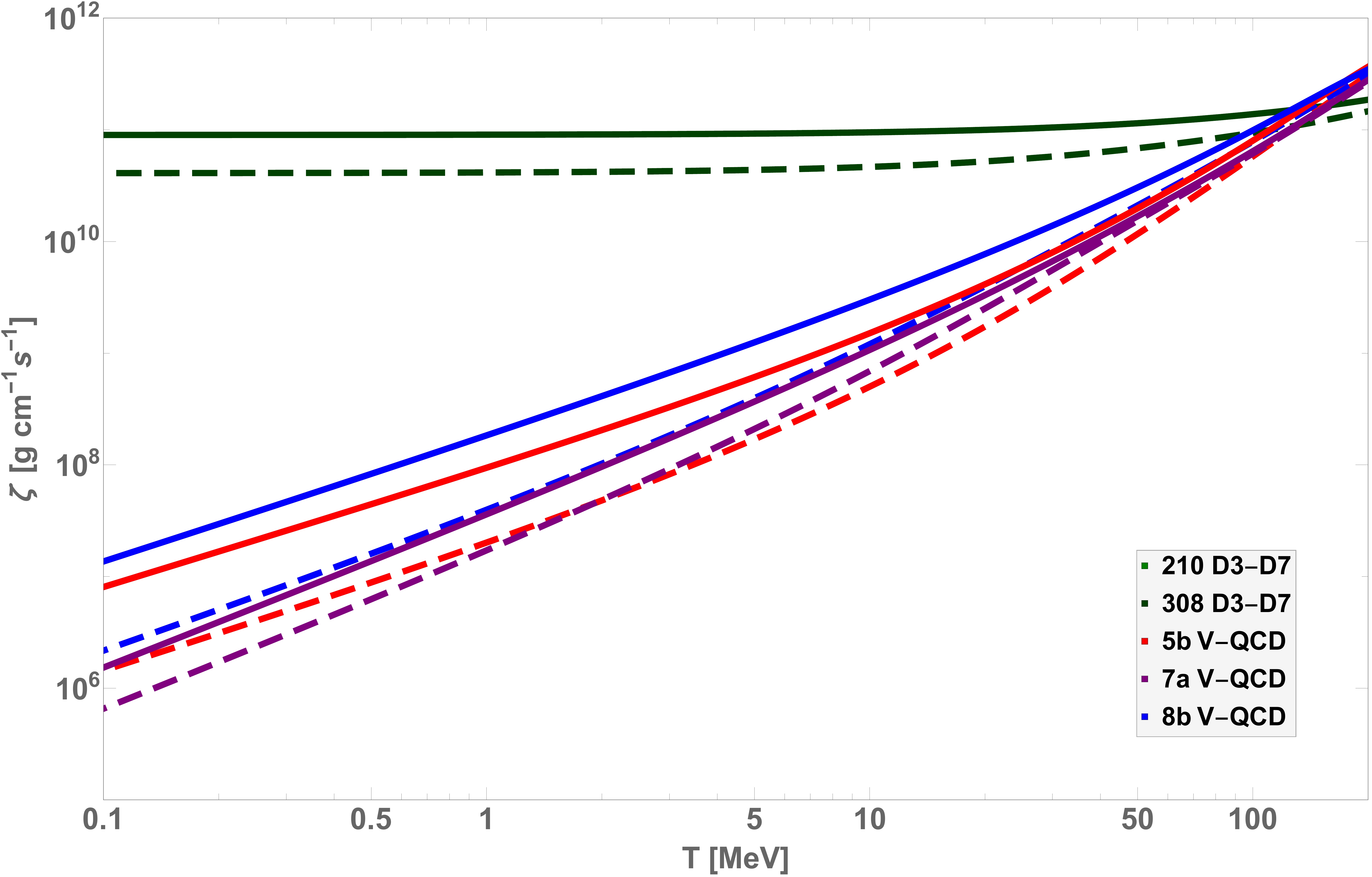}
    \caption{\small Shear (right) and bulk (left) viscosities from holography (curves) compared to perturbation theory (green band). The solid (dashed) curves are at $\mu=450$~MeV (600~MeV). Figures from~\cite{Hoyos:2020hmq,Hoyos:2021njg}.}
    \label{fig:viscosities}
\end{figure*}

We computed estimates for transport properties using the D3-D7 model and V-QCD in~\cite{Hoyos:2020hmq,Hoyos:2021njg}, and compared to perturbation theory. The results from holography were seen to deviate significantly from those obtained from perturbation theory. This suggests that one should be cautious when trying to estimate neutron star transport based on any perturbative analysis. Our result for the shear (left) and bulk (right) viscosities are shown in Fig.~\ref{fig:viscosities}.

\subsection{The holographic baryon}\label{sec:baryon}

As commented above, the precise way of introducing nucleons in holographic models is via soliton configurations of non-Abelian gauge/fields in the bulk, which arise from the DBI actions~\cite{Witten:1998xy,Gross:1998gk}. Such constructions have been carried out, for example, in the Witten-Sakai-Sugimoto model~\cite{Kim:2006gp,Hata:2007mb,Bolognesi:2013nja} and in the hard wall bottom-up model~\cite{Pomarol:2007kr,Pomarol:2008aa}. The coupling of the soliton to the tachyon field, which characterizes chiral symmetry breaking, was studied in~\cite{Gorsky:2013dda,Gorsky:2015pra}. The setup for the baryon solution in V-QCD was established in~\cite{Jarvinen:2022mys}. This required, among other things, a thorough study of the possible Chern-Simons action in the bottom-up framework, in the presence of nontrivial tachyon field. Numerical results for the soliton will appear soon~\cite{Jarvinen:2022xxx}. After fitting the model parameters to thermodynamics, meson masses, and the pion decay constant, the classical mass of the soliton was found to be near $1$~GeV, in good agreement with the observed masses of the nucleons.

\subsection{Equation of state at finite temperature} 

State-of-the-art neutron star merger simulations require EOSs which depend, apart from density, the temperature and the charge fraction. The temperature dependence is important because shocks created during the merger heat up the matter significantly, and the charge fraction is important because the rapid merger process drives the matter outside $\beta$-equilibrium. 

Consequently, the cold hybrid V-QCD EOS of~\cite{Ecker:2019xrw,Jokela:2020piw} was extended to finite $T$ and outside of $\beta$-equilibrium in~\cite{Demircik:2021zll}. A key ingredient in this construction was a van-der-Waals model 
(see, e.g.,~\cite{Vovchenko:2017zpj}) 
which was adjusted such that it 
agrees exactly with the V-QCD EOS at $T=0$. 
It therefore provides an extrapolation of the V-QCD dense nuclear matter EOS to finite $T$. At low densities, and for the dependence on the charge fraction, we used the HS(DD2) model~\cite{Hempel:2009mc,Typel:2009sy}, 
another key ingredient in this construction. 
Since both the quark matter and nuclear matter phases are described by V-QCD (apart from mild temperature dependence in the nuclear matter phase coming from the van-der-Waals extrapolation), the phase transition and the nuclear-quark matter mixed phase are determined by a single model, V-QCD. 

The EOS was constructed in three variants, which roughly represent the leftover freedom in the parameters of V-QCD after fitting the model to lattice data.\footnote{The EOS files  can be downloaded from the CompOSE database~\cite{Typel:2013rza,Typel:2022lcx} for the three variants: \href{https://compose.obspm.fr/eos/290}{DEJ(DD2-VQCD) soft}, \href{https://compose.obspm.fr/eos/289}{DEJ(DD2-VQCD) intermediate}, \href{https://compose.obspm.fr/eos/291}{DEJ(DD2-VQCD) stiff}. 
} In each of these models, the nuclear to quark matter transition was seen to end in a critical point with
\begin{equation}
 110~\mathrm{MeV} \lesssim T_c \lesssim 130~\mathrm{MeV}, \qquad  480~\mathrm{MeV} \lesssim \mu_{bc} \lesssim 580~\mathrm{MeV} \ ,
\end{equation}
which are close to estimates obtained in simpler holographic models~\cite{DeWolfe:2010he,Knaute:2017opk,Critelli:2017oub,Cai:2022omk}.

\subsection{Quark matter production in neutron star mergers}

The finite temperature V-QCD EOSs were used in state-of-the-art neutron star merger simulations in~\cite{Tootle:2022pvd}. That is, we evolved four dimensional general relativity and hydrodynamics by using the holographic EOS as an input. The chirp mass for the mergers was chosen to match with the GW170817 event, and we ran simulations with different mass ratios. 
The simulations used the Frankfurt University/Kadath (FUKA)~\cite{Papenfort:2021hod} for initial data and the Frankfurt/Illinois (FIL) code~\cite{Most:2019kfe} for evolution.

We focused on the production of quark matter in the merger. With the V-QCD EOSs, isolated neutron stars are fully composed of nuclear matter, but quark matter can be generated during the merger process, in particular right before a potential collapse into a black hole. We classified quark matter production into three stages: hot, warm and cold quark matter. 
In the first stage, hot quark matter is produces in the hottest (but not densest) regions of the hypermassive neutron star right after the merger. Warm quark matter is produced later due to complex merger dynamics in regions that are neither hottest nor densest. Finally, cold quark matter is produced in the dense and cold core right before the collapse to a black hole. 

\section{Conclusion}
I discussed how the gauge/gravity duality, in combination with other methods and data from QCD, can be used to constrain the physics of dense QCD matter and to derive new results. I focused on the V-QCD model, with which many details work nicely. In particular I obtained
\begin{itemize}
 \item A feasible model for quark matter, including a precise fit of lattice QCD thermodynamics and  reasonable extrapolation to the region of dense and cold matter.
 \item Nuclear matter model with stiff EOS, as preferred by astrophysical observations.
 \item Simultaneous description of nuclear and quark matter (including the transition between them and its critical end point) in a single framework.
\end{itemize}

Future work will extend the model by including dependence on quark flavors (in particular the strange quark mass) and possibly paired, color superconducting phases. Another direction is to study the coupling of the strongly interacting matter to weak interactions. 

\section*{Acknowledgments}
MJ has been
supported by an appointment to the JRG Program at the APCTP through the
Science and Technology Promotion Fund and Lottery Fund of the Korean
Government. MJ has also been supported by the Korean Local
Governments -- Gyeong\-sang\-buk-do Province and Pohang City -- and by
the National Research Foundation of Korea (NRF) funded by the Korean
government (MSIT) (grant number 2021R1A2C1010834). 
\bibliography{mj-biblio.bib}

\end{document}